\documentclass[twocolumn,aps,pre,longbibliography]{revtex4-1}
\usepackage{amsmath,amssymb}
\usepackage{graphicx}% Include figure files
\usepackage{dcolumn}% Align table columns on decimal point
\usepackage{color}
\usepackage{epstopdf}
\usepackage[%
colorlinks=true,
urlcolor=blue,
linkcolor=blue,
citecolor=blue
]{hyperref}
\begin{document}
	\hyphenpenalty=5000
	%\tolerance=1000
	\hyphenation{ma-the-ma-ti-cal equi-li-brium sub-system tem-pera-ture eigen-levels distri-bution evo-lu-tion pro-ba-bi-li-ty tra-jectory macro-scopic micro-scopic quan-tum pro-perty ca-no-ni-cal sub-systems des-crip-tion tem-pe-ra-ture thermo-dynamics de-ge-ne-ra-cy thermo-dynamic phe-no-me-na phe-no-me-no-lo-gi-cal appli-cation theory Neumann eigen-energy react-ant ca-te-gor-iza-tion dif-ferent dif-ference patterns}
\title{Steepest-entropy-ascent quantum thermodynamic modeling of heat and mass diffusion in a far-from-equilibrium system based on a single particle ensemble}
\author{Guanchen Li}
\email{guanchen@vt.edu}
\author{Michael R. von Spakovsky}
\email{vonspako@vt.edu}
\affiliation{%
	Center for Energy Systems Research, Mechanical Engineering Department\\
	Virginia Tech, Blacksburg, VA 24061
}%
\date{\today}
\begin{abstract}
This paper presents a nonequilibrium thermodynamic model for the relaxation of a local, isolated system in nonequilibrium using the principle of steepest entropy ascent (SEA), which can be expressed as a variational principle in thermodynamic state space. The model is able to arrive at the Onsager relations for such a system. Since no assumption of local equilibrium is made, the conjugate fluxes and forces, which result, are intrinsic to the subspaces of the system's state space and are defined using the concepts of hypoequilibrium state and nonequilibrium intensive properties, which describe the non-mutual equilibrium status between subspaces of the thermodynamic state space. The Onsager relations are shown to be a thermodynamic kinematic feature of the system independent of the specific details of the micro-mechanical dynamics. Two kinds of relaxation processes are studied with different constraints (i.e., conservation laws) corresponding to heat and mass diffusion. Linear behavior in the near-equilibrium region as well as nonlinear behavior in the far-from-equilibrium region are discussed. Thermodynamic relations in the equilibrium and near-equilibrium realm, including the Gibbs relation, the Clausius inequality, and the Onsager relations, are generalized to the far-from-equilibrium realm. The variational principle in the space spanned by the intrinsic conjugate fluxes and forces is expressed via the quadratic dissipation potential. As an application, the model is applied to the heat and mass diffusion of a system represented by a single particle ensemble, which can also be applied to a simple system of many particles. Phenomenological transport coefficients are also derived in near-equilibrium realm.
\end{abstract}
\maketitle
\section{Introduction}
The study of nonequilibrium relaxation processes including chemical kinetics, mass diffusion, and heat diffusion is typically accomplished using approaches based on microscopic mechanics or thermodynamics. With the former, system state space is spanned by microstates and the governing equation is based on the dynamics of classical mechanics (e.g., molecule dynamics \cite{Rapaport2004} or kinetic theory \cite{Chen1998}), quantum mechanics (e.g., nonequilibrium Green's functions \cite{Vogl2010} or the quantum Boltzmann equation, i.e., the Uehling-Unlenbeck-Boltzmann equation \cite{Rossani2000,Garcia2003,Torres-Rincon2014}) or a stochastic process (e.g., Monte Carlo simulation of the Ising model \cite{Newman1999}). These approaches provide complete information of the microscopic process such as individual particle collisions or quantum state scattering. However, the large amount of very detailed information required inevitably results in large computational burdens, which limit the applicability of these approaches.

Approaches based on thermodynamics, on the other hand, are not similarly burdened and are able to generally capture the features of the relaxation process is of interest via, for example, the Onsager relation. Approaches of this type include nonequilibrium thermodynamics \cite{Groot1962,Gyarmati1970}, linear response functions and the fluctuation-dissipation theorem \cite{Kubo2012}, stochastic thermodynamics \cite{Seifert2012}, extended irreversible thermodynamics \cite{Jou1996}, etc. The thermodynamic features captured can be regarded as a coarse graining of the microscopic dynamics or as a pattern in ensemble evolution \cite{Grmela1997,Ottinger1997}, which computationally is more efficient. However, most of these approaches have limited or no applicability in the far-from-equilibrium realm, since the local or near-equilibrium assumption is needed or analytical solutions are only available at steady state. In addition, their governing equations are phenomenological or stochastic in nature and, thus, do not have a first-principle basis. To address these issues and push the application of thermodynamic principles further into the nonequilibrium realm, it is of great importance to find a general and simple description of nonequilibrium state corresponding to a thermodynamic pattern of the microscopic description, to fundamentally define the macroscopic properties of any thermodynamic state (i.e., extensive or intensive properties for both equilibrium and nonequilibrium states), and to use a thermodynamic governing equation based on first principles. 

Steepest-entropy-ascent quantum thermodynamics (SEAQT), which is a first-principle, thermodynamic-ensemble based approach, addresses all of the issues raised above, providing a governing equation able to describe the nonequilibrium process from an entropy generation viewpoint. The description of system state is based on the density operator of quantum mechanics or probability distribution in state space of classical mechanics. The macroscopic properties of entropy, energy, and particle number, which are well defined for any state of any system \cite{Zanchini2014}, are used to develop the governing equation and describe system state evolution. Recently, this description has been simplified via the concept of hypoequilibrium state \cite{LivonSpakovsky2015}, which captures the global features of the microscopic description for the relaxation process. In addition, the concept of nonequilibrium intensive properties introduced in \cite{LivonSpakovsky2015} based on the concept of hypoequilibrium state enables a complete description of the nonequilibrium evolution of state when combined with the set of nonequilibrium extensive properties. Unlike the intensive property definitions of other nonequilibrium thermodynamic approaches (definitions which require local-equilibrium, near-equilibrium, or steady state assumption or a phenomenological basis), the definitions in the SEAQT framework are fundamental and available to all nonequilibrium states and are especially suitable for the description of the evolution in state of relaxation processes. Both of these concepts enables the generalization of many equilibrium (or near-equilibrium) thermodynamic relations such as the Gibb's relation, the Clausius inequality, and the Onsager relations into the far-from-equilibrium realm.

To describe the SEAQT framework, the paper starts with a derivation in Secs. II.A and B of the SEAQT relaxation dynamics starting from the geometry of system state space. Some useful mathematical features of the relaxation process are then presented in Sec. II.C, enabling the definition of the concepts of hypoequilibrium state and nonequilibrium intensive properties. As an alternative to the geometric derivation from state space, the relaxation dynamics of SEAQT can also be derived using a variational principle in system state space as is done in Sec. II.D. Section III then follows with a discussion of mass diffusion in a local, isolated system in nonequilibrium. Subsequently, the transport equations, the Gibbs relation, and the Onsager relations are generalized for the nonequilibrium relaxation process, and the dissipation potential is given to complete the link from the variational principle in system state space to that for the conjugate fluxes and/or forces \cite{Gyarmati1970}. Section IV discusses the process of heat diffusion by choosing a set of system constraints different from those used for mass diffusion. In Sec. V, the SEAQT model is applied to the study of the heat and mass diffusion of a simple system consisting of ideal gas (hydrogen), which can be represented by a single-particle ensemble. Linear behavior in the near-equilibrium realm provides the phenomenological transport equations, and higher-order, nonlinear behavior in the near-equilibrium realm is also studied. Finally, section VI concludes the paper with some final comments.

\section{Theory: SEA-QT Equation of motion}
The basic framework of SEAQT is introduced in this section. This framework describes the relaxation process of a local, isolated system in nonequilibrium based on thermodynamic concepts. To begin with, the equation of motion for such a system is derived in Sec. II.A from the geometric principle of steepest entropy ascent. This is followed in Sec. II.B by a discussion of the kinetics and dynamics of the relaxation process, which enable the study of the thermodynamic trajectory regardless of the details of the microscopic interactions. A description of nonequilibrium state and its evolution for the relaxation process is then given in Sec. II.C using the concepts of hypoequilibrium state and nonequilibrium intensive properties. Finally, in Sec. II.D, a presentation of the variational principle of steepest entropy ascent is used to derive the SEAQT equation of motion for the local system.

\subsection{SEAQT equation of motion for an isolated system}
Based on the discussion by Grmela \cite{Grmela2014,Grmela1997,Ottinger1997} and Beretta \cite{Beretta2014,Montefusco2015}, the general form of a nonequilibrium framework is a combination of both irreversible relaxation and reversible symplectic dynamics. If written in the generalized form of the Ginzburg-Landau equation \cite{Grmela1997,Montefusco2015}, the equation of motion takes the following form:
\begin{equation}
\frac{d}{dt}\gamma(t)=X^{H}_{\gamma(t)}+Y^{H}_{\gamma(t)}
\end{equation}
where $\gamma(t)$ represents the state evolution trajectory, $X^{H}_{\gamma(t)}$ and $Y^{H}_{\gamma(t)}$ are functions of the system state $\gamma(t)$ and represent the reversible symplectic dynamics and irreversible relaxation dynamics, respectively. In the SEAQT framework, the system state is represented by the density operator $\hat{\rho}$, $X^{H}_{\gamma(t)}$ follows the Schr\"{o}dinger equation, and $Y^{H}_{\gamma(t)}$ is derived from the SEA principle. Thus,
\begin{equation}
\frac{d\hat{\rho}}{dt}=\frac{1}{i\hbar}[\hat{\rho},\hat{H}]+\frac{1}{\tau(\hat{\rho})}\hat{D}(\hat{\rho})
\end{equation}
where $\hat{D}$ is the dissipation operator determined via a constrained gradient in Hilbert space. A metric tensor must be specified in the derivation of this dissipation term, since it describes the geometric features of the Hilbert space \cite{Beretta2014}. $\tau$ is the relaxation time, which represents the speed of system evolution in Hilbert space.

In the application here to the modeling of heat and mass diffusion, the system is restricted to the class of dilute-Boltzmann-gas states in which the particles have no quantum correlation between eigenstates, and are independently distributed \cite{Beretta2006}. Such states can be represented by a single-particle density operator that is diagonal in the basis of the single-particle energy eigenstates. The Hilbert space metric chosen here is the Fisher-Rao metric, which is uniform in different dimensions of Hilbert space. Under these conditions, the symplectic Sch\"{o}dinger term in the equation of motion vanishes. Thus, the focus is this paper is on the irreversible relaxation process only.

A group of energy eigenlevels $\{\epsilon_i, i=1,2,...\}$ is determined from the system Hamiltonian. The state of the system can then be represented by a probability distribution among the energy eigenlevels $\{p_i, i=1,2,...\}$, which is the diagonal terms of the density operator. Using the Fisher-Rao metric of the probability space $\{p_i, i=1,2,...\}$, one can define the distance in probability space, which can be used as the state distance. Equivalently, the square root of the probability distribution $\{x_i, i=1,2,...\}$ can be used to represent the system state with the result that the Fisher-Rao metric of the probability space becomes the Euclidean metric in square root of the probability space of $\{x_i, i=1,2,...\}$. The later representation is used in the paper. Both are expressed as follows:
\begin{eqnarray}
&&\text{State: } \{p_i, i=1,2,...\},\nonumber\\
&&\text{Distance: }  dl = \frac{1}{2}\sqrt{\sum_i p_i(\frac{d\ln p_i}{d\theta})^2}d\theta\\
&&\text{State: } \{x_i, i=1,2,...\},\nonumber\\
&&\text{Distance: }  dl = \sqrt{\sum_i (dx_i)^2}
\end{eqnarray}
where $dl$ is the distance between $p(\theta+d\theta)$ and $p(\theta)$ or $x(\theta+d\theta)$ and $x(\theta)$, and $\theta$ is a continuous parameter. An extensive property of the system can then be defined as a function of the state $\{x_i\}$ such that
\begin{eqnarray}
I & = & \sum_{i} x_{i}^2\label{5}\\
E & = &\langle e\rangle =\sum_{i} \epsilon_i x_{i}^2\label{6}\\
S & = &\langle s\rangle = \sum_{i} -x_{i}^2 \ln({x_{i}^2})\label{7}
\end{eqnarray}
where $\langle\dots\rangle$ means the ensemble average. The von Neumann formula for entropy is used, because as shown in \cite{Gyftopoulos1997}, it has all the properties required by thermodynamics. The gradient of a given property in state space is then expressed by
\begin{eqnarray}
\boldsymbol{g_I} & = & \sum_{i}\frac{\partial I}{\partial x_i}\hat{e_{i}} = \sum_{i} 2x_{i}\hat{e_{i}}\\
\boldsymbol{g_E} & = & \sum_{i}\frac{\partial E}{\partial x_i}\hat{e_{i}} = \sum_{i} 2\epsilon_i x_{i}\hat{e_{i}}\\
\boldsymbol{g_S} & = & \sum_{i}\frac{\partial S}{\partial x_i}\hat{e_{i}} = \sum_{i} [-2x_{i}-2x_{i} \ln(x_i^2)]\hat{e_{i}}
\end{eqnarray}
where $\hat{e_{i}}$ is the unit vector for each dimension. Furthermore, for an isolated system, the system satisfies the conservation laws for probability and energy, i.e.,
\begin{equation}
I = \sum_{i} x_{i}^2= 1,\quad E = \sum_{i} \epsilon_i x_{i}^2=\text{constant}
\end{equation}
The principle of SEA upon which the equation of motion is based is defined as the system state evolving along the direction that at any instant of time has the largest entropy gradient consistent with the conservation constraints. The equation of motion is given by
\begin{equation}
\frac{d\boldsymbol{x}}{dt} =  \frac{1}{\tau(x)}\boldsymbol{g_S}_{\perp L(\boldsymbol{g_I},\boldsymbol{g_E})}
\end{equation}
where $\tau$, which is a function of system state, is the relaxation time which describes the speed at which the state evolves in state space in the direction of steepest entropy ascent. $L(\boldsymbol{g_I},\boldsymbol{g_E})$ is the manifold spanned by $\boldsymbol{g_I}$ and $\boldsymbol{g_E}$, and $\boldsymbol{g_S}_{\perp L(\boldsymbol{g_I},\boldsymbol{g_E})}$ is the perpendicular component of the gradient of the entropy to the hyper-surface that conserves the probability and energy. It takes the form of a ratio of Gram determinants expressed as
\begin{equation}\label{Gram}
\boldsymbol{g_S}_{\perp L(\boldsymbol{g_I},\boldsymbol{g_E})}=\frac{\left|\begin{array}{ccc}
\boldsymbol{g_S} & \boldsymbol{g_I} & \boldsymbol{g_E} \\
(\boldsymbol{g_S},\boldsymbol{g_I}) & (\boldsymbol{g_I},\boldsymbol{g_I}) & (\boldsymbol{g_E},\boldsymbol{g_I}) \\
(\boldsymbol{g_S},\boldsymbol{g_E}) & (\boldsymbol{g_I},\boldsymbol{g_E}) & (\boldsymbol{g_E},\boldsymbol{g_E})
\end{array}\right|}{\left|\begin{array}{cc}
(\boldsymbol{g_I},\boldsymbol{g_I}) & (\boldsymbol{g_E},\boldsymbol{g_I}) \\
(\boldsymbol{g_I},\boldsymbol{g_E}) & (\boldsymbol{g_E},\boldsymbol{g_E}) 
\end{array}\right|}
\end{equation}
where $(\dots,\dots)$ denotes the scalar product of two vectors in state space. The explicit form of Eq. (\ref{Gram}) for $\{p_j\}$ is, thus, \cite{Beretta2006}
\begin{equation}\label{SEAQTEM}
\frac{dp_j}{dt}=\frac{1}{\tau}\frac{\left|\begin{array}{ccc}
-p_j\ln p_j & p_j & \epsilon_jp_j \\
\langle s\rangle & 1 & \langle e\rangle \\
\langle es\rangle & \langle e\rangle & \langle e^2\rangle
\end{array}\right|}{\left|\begin{array}{cc}
1 & \langle e\rangle \\
\langle e\rangle & \langle e^2\rangle 
\end{array}\right|}
\end{equation}
where
\begin{equation}
\langle e^2\rangle = \sum_{i} \epsilon_i^2 x_{i}^2, \quad \langle es\rangle = \sum_{i} -\epsilon_ix_{i}^2 \ln(x_{i}^2)
\end{equation}
The state representation and the equation of motion can be simplified by combining degenerate energy eigenlevels \cite{LivonSpakovsky2015}. The system is defined by a group of different energy eigenlevels $\{\epsilon_i, i=1,2,...\}$ and their degeneracy $\{n_i, i=1,2,...\}$. The state of the system is described by a probability distribution among the energy eigenlevels $\{p_i, i=1,2,...\}$ or square root of the probability $\{x_i, i=1,2,...\}$. As a result, the equation of motion changes to
\begin{equation}
\frac{dp_j}{dt}=\frac{1}{\tau}\frac{\left|\begin{array}{ccc}
-p_j\ln \frac{p_j}{n_j} & p_j & \epsilon_jp_j \\
\langle s\rangle & 1 & \langle e\rangle \\
\langle es\rangle & \langle e\rangle & \langle e^2\rangle
\end{array}\right|}{\left|\begin{array}{cc}
1 & \langle e\rangle \\
\langle e\rangle & \langle e^2\rangle 
\end{array}\right|}
\end{equation}
where the properties are defined by
\begin{eqnarray}
\langle e\rangle =\sum_{i} \epsilon_i x_{i}^2&,& \quad\langle s\rangle = \sum_{i} -x_{i}^2 \ln(\frac{{x_{i}^2}}{n_i})\nonumber\\
\langle e^2\rangle = \sum_{i} \epsilon_i^2 x_{i}^2&,& \quad \langle es\rangle = \sum_{i} -\epsilon_ix_{i}^2 \ln(\frac{{x_{i}^2}}{n_i})
\end{eqnarray}

\subsection{Nonequilibrium evolution: Kinetics and Dynamics}
In general, the equation of motion for a system with a given group of conservation laws has the form:
\begin{equation}\label{GEM}
\frac{dp_j}{dt}=\frac{1}{\tau(\mathbf{p})}D_j(\mathbf{p})
\end{equation}
where $D_j(\mathbf{p})$ is calculated from the conservation laws and the principle of steepest entropy ascent \cite{Beretta2009}. Specifically, it takes the form of Eq. (\ref{SEAQTEM}) for an isolate system yielding to mass and energy conservation.

Since for a given initial state of the system, the nonequilibrium thermodynamic path of state evolution is uniquely solved from this equation, Eq. (\ref{GEM}), the path can be used to define a new parameter $\tilde{\tau}$ given by
\begin{equation}\label{tau_change}
d\tilde{\tau}=\frac{1}{\tau(\mathbf{p}(t))}dt, \quad\text{or}\quad \tilde{\tau}=\int_{path}\frac{1}{\tau(\mathbf{p}(t'))}dt'=\tilde{\tau}(t)
\end{equation}
where $\tilde{\tau}$ is called the dimensionless time. With this time, the independent variable for the equation of motion can be changed so that
\begin{equation}\label{GEM_K}
\frac{dp_j}{d\tilde{\tau}}=D_j(\mathbf{p})
\end{equation}
The solution of this equation is written as:
\begin{equation}\label{GEM_K_tau}
p_j=p_j(\tilde{\tau})
\end{equation}
Independent of how the relaxation time $\tau$ depends on the real time $t$ and the state, the equation of motion can always be transformed to Eq. (\ref{GEM_K}) with the parameter change defined by Eq. (\ref{tau_change}). Furthermore, the evolution of system state will follow the same function [Eq. (\ref{GEM_K_tau})] in $\tilde{\tau}$. Physically, this means that the system follows the same trajectory in state space. By using this transformation, the kinetics and dynamics of the system are separated. The former is found via Eqs. (\ref{GEM_K}) and (\ref{GEM_K_tau}) and result in the trajectory in state space based on the parameter $\tilde{\tau}$ or a constant relaxation time $\tau$. The dynamics are found via Eq. (\ref{GEM}) and the functional dependence $\tau=\tau(p)$ [Eq. (\ref{tau_change})] and result in the trajectory in state space based on the real time $t$.

In the discussion on mass diffusion (Sec. III) and heat diffusion (Sec. IV), it is shown that the kinetics (or its associated trajectory) of the nonequilibrium relaxation results in a generalized Gibbs relation and the Onsager relations in the far-from-equilibrium realm and in linear phenomenological equations in the near-equilibrium realm, which are independent of the dynamics. The kinetics appears as a system feature or pattern of the thermodynamics in the sense of the GENERIC \cite{Grmela1997,Ottinger1997,Grmela2014}, which is an ensemble or group feature. Information about the mechanical details (e.g., how the particles interact in the system mechanically) can be included in the dynamics (e.g., by how $\tau$ is chosen) when the state evolution in time $t$ is studied. The focus of this paper, however, is on the thermodynamic features of the nonequilibrium relaxation so that $\tau$ is set equal to $1$. More discussion on how $\tau$ is chosen using the mechanics and on the dynamics of nonequilibrium is left for a future paper.

\subsection{Nonequilibrium state and state evolution description: Hypoequilibrium}
The solution of the SEA-QT equation of motion exhibits some good properties, which allows for a complete description of nonequilibrium state and the general definition of nonequilibrium temperature. More discussion is presented in reference \cite{LivonSpakovsky2015}, and an example is provided below. The energy eigenlevels of the system $\{\epsilon_i,i=1,2,3...\}$ with degeneracy $\{n_i,i=1,2,3...\}$ can be divided into to $M$ sets $\{\epsilon_i^K\}$ (degeneracy $\{n_i^K\}$) with $i=1,2,3,...,\,K=1,2,...,M$, so that the state space of the system (the Hilbert space) $\mathcal{H}$ can be represented by the sum of $M$ subspaces $\mathcal{H}_K$, with $K=1,2,...,M$, i.e.,
\begin{equation}
\mathcal{H}=\bigoplus_{K=1}^M\mathcal{H}_K
\end{equation}
To be complete, $M$ can be infinite. The system state can be represented by the distributions in $M$ subspace energy eigenlevels $\{p_i^K, K=1,...,M\}$. If the probability distribution in one subspace, for example, the $K$th subspace yields to the canonical distribution of parameter $\beta_K$, the temperature of the $K$th subspace is defined to be $T_K=1/\beta_K$. Given a way to divide the energy eigenlevels, if the system probability distributions in the $M$ subspaces are all canonical distribution, the state of the system is called a $M$th-order hypoequilibrium state \cite{LivonSpakovsky2015}, which can be described uniquely by the total probability in each subspace ($\{p^K=\sum p_i^K, K=1,...,M\}$) and the temperature of each of the subspaces ($\{T_K,K=1,...,M\}$). If the initial state of the system is a $M$th-order hypoequilibrium state then
\begin{eqnarray}\label{HEPEi}
p_i^K(t=0)&=&\frac{p^Kn_i^K}{Z^K (\beta^K)} e^{-\beta_K\epsilon_i^K},i=1,2,3,...
\end{eqnarray}
where $k_b$ is the Boltzmann constant, $Z^K (\beta^K)$ is the partition function of subspace $K$ at temperature $T^K$. A more general form to represent any nonequilibrium state is given in \cite{Beretta2009} using the language of quantum mechanics. Li and von Spakovsky \cite{LivonSpakovsky2015} have proven that the system retains a $M$th-order hypoequilibrium state throughout the nonequilibrium relaxation process if it initial starts out in such a state. The solution to Eq. (\ref{GEM}), thus, becomes   
\begin{equation}\label{HEPE}
p_i^K(t)=\frac{p^K(t)}{Z^K (\beta^K(t))} n_i^Ke^{-\beta^K(t)\epsilon_i^K},i=1,2,3,...
\end{equation}
As a result, each subspace has temperature defined throughout the entire nonequilibrium relaxation process. This result applies to an isolate system with probability and energy conservation. For a system with a different set of conservation laws, a similar relation exists. However, the general proof is left for a future paper. The proof for a system with heat diffusion only is given in the Appendix A.

\subsection{Local variational principle in thermodynamic state space}
According to Beretta \cite{Beretta2006}, the equation of motion can be derived from a local variational principle, which can be regarded as the variational form of the steepest entropy ascent principle, i.e.,
\begin{eqnarray}
&&\text{Maximize: }\dot{S}(\dot{\mathbf{x}})=(\dot{\mathbf{x}},\mathbf{g_S})\,\,\text{subject to }\nonumber\\ &&\dot{E}=(\dot{\mathbf{x}},\mathbf{g_E})=0,\,\dot{I}=(\dot{\mathbf{x}},\mathbf{g_I})=0,\,(\dot{\mathbf{x}},\,\dot{\mathbf{x}})=\xi(\mathbf{x})\nonumber\\
&&\text{with } \delta \dot{\mathbf{x}}\neq 0,\,\delta \mathbf{x}= 0
\end{eqnarray}
The third constraint on $\dot{\mathbf{x}}$ indicates that only the direction of $\dot{\mathbf{x}}$ is of interest. This variational principle is in microscopic state space, which contrasts with the variational principle in the space spanned by conjugate fluxes and forces presented later for the Onsager relations.

\section{Theory: Diffusion in a nonequilibrium system}

In the next two sections, Secs. III and IV, two interactions are studied with the SEAQT framework. The theory for mass diffusion in a local, isolated system in nonequilibrium is presented in this section, Sec. III. The Gibbs relation, the entropy generation for a non-quasi-equilibrium process, and the Onsager relations are derived based on the concept of hypoequilibrium state and intensive properties. The variational principle using conjugate forces is given at the end of Sec. III.
\subsection{Equation of motion for mass diffusion}
The mass (or probability) diffusion across energy eigenlevels (or across subspaces) can be studied for an isolated system in nonequilibrium. Using Eq. (\ref{SEAQTEM}), one energy eigenlevel in the $K$th subspace yields to the following equation of motion: 
\begin{equation} \label{EQM}
\frac{dp_j^K}{dt}=\frac{1}{\tau}(-p_j^K\ln \frac{p_j^K}{n_j^K}-p_j^K\frac{A_2}{A_1}+\epsilon_jp_j^K\frac{A_3}{A_1})
\end{equation}
where
\begin{eqnarray}
A_1=\left|\begin{array}{cc}
1 & \langle e\rangle \\
\langle e\rangle & \langle e^2\rangle 
\end{array}\right|,
A_2=\left|\begin{array}{cc}
\langle s\rangle & \langle e\rangle \\
\langle es\rangle & \langle e^2\rangle 
\end{array}\right|,
A_3=\left|\begin{array}{cc}
\langle s\rangle & 1 \\
\langle es\rangle & \langle e\rangle 
\end{array}\right|\nonumber\\
\end{eqnarray}
Summation over all energy eigenlevels in the $K^{th} $subspace yields the evolution of the probability in the $K$th subspace, represented by $p^K$, namely,
\begin{equation}\label{28}
\frac{dp^K}{dt}=\frac{1}{\tau}(-p^K\ln p^K+p^K\langle \tilde{s}\rangle^K -p^K\frac{A_2}{A_1}+p^K\langle \tilde{e}\rangle^K\frac{A_3}{A_1})
\end{equation}
where $\langle \tilde{\dots}\rangle^K$ is the specific property in the $K$th subspace. To calculate the specific property, the probability distribution in the $K$th subspace is found from
\begin{equation}
p^K\equiv\sum_j p_j^K,\quad \tilde{p}_j^K\equiv\frac{p_j^K}{p^K}
\end{equation}
where $p^K$ is the particle number in the $K$th subspace. The specific property in the $K$th subspace is then expressed as
\begin{eqnarray}
&&\langle \tilde{e}\rangle^K \equiv \sum_j \epsilon_j^K\tilde{p}_j^K\label{eK}\\
&&\langle \tilde{s}\rangle^K \equiv -\sum_j \tilde{p}_j^K\ln \frac{\tilde{p}_j^K}{n_j^K}\label{sK}
\end{eqnarray}
%=\sum_j \tilde{p}_j^K\tilde{s}_j^K,\,\tilde{s}_j^K \equiv -\ln \frac{\tilde{p}_j^K}{n_j^K}
%\end{eqnarray}
%where $\tilde{s}_j^K$ is the entropy contribution from one 
\subsection{Particle number and temperature evolution when the initial state is a hypoequilibrium state}

If the system is in a $M$th-order hypoequilibrium state, the probability evolution yields Eq. (\ref{HEPE}). For simplicity, the following definition is made:
\begin{equation}
\alpha^K = \ln Z^K-\ln p^K
\end{equation}
With this definition, the probability evolution of one energy eigenlevel is given by
\begin{eqnarray}\label{HES}
p_i^K(t)&=&\frac{p^K(t)}{Z^K (\beta^K(t))} n_i^Ke^{-\beta^K(t)\epsilon_i^K}\nonumber\\
&=&n_i^Ke^{-\alpha^K(t)-\beta^K(t)\epsilon_i^K}
\end{eqnarray}
$\alpha^K$ and $\beta^K$ are nonequilibrium intensive properties of the $K$th subspace, corresponding to the extensive properties $p^K$ and $E^K\equiv p^k\langle \tilde{e}\rangle^K$. Furthermore, by defining
\begin{equation}\label{34}
\alpha = \frac{A_2}{A_1}, \quad \beta = -\frac{A_3}{A_1}
\end{equation}
the particle number and energy evolution of the $K$th subspace can be acquired from Eq. (\ref{EQM}), i.e.,
\begin{eqnarray}\label{PE}
\frac{dp^K}{dt}&=&\frac{1}{\tau}p^K(\alpha^K-\alpha)+\frac{1}{\tau}p^K \langle \tilde{e}\rangle^K(\beta^K-\beta)\\
\frac{d\langle e\rangle^K}{dt}&=&\frac{1}{\tau}p^K\langle \tilde{e}\rangle^K(\alpha^K-\alpha)+\frac{1}{\tau}p^K \langle \tilde{e^2}\rangle^K(\beta^K-\beta)\quad\label{EE}
\end{eqnarray}
From Eqs. (\ref{HEPE}) and (\ref{EQM}), the intensive properties $\alpha^K$ and $\beta^K$ obey the evolutions (see Appendix B for the derivation)
\begin{eqnarray}
\frac{d\alpha^K}{dt}&=&-\frac{1}{\tau}(\alpha^K-\alpha)\label{aE}\\
\frac{d\beta^K}{dt}&=&-\frac{1}{\tau}(\beta^K-\beta)\label{bE}
\end{eqnarray}
The authors prove that $\alpha$ and $\beta$ have the physical meaning of intensive properties from measurements of a nonequilibrium state \cite{LivonSpakovsky2015b}. At stable equilibrium, the intensive properties in any subsystem obey the following relations:
\begin{eqnarray}
\alpha(t=t^{eq}) = \alpha^K(t=t^{eq}) =\alpha^{eq}\\
\beta(t=t^{eq}) = \beta^K(t=t^{eq}) = \beta^{eq}
\end{eqnarray}

\subsection{Gibbs relation, entropy generation for a non-quasi-equilibrium process, and the Onsager relations in the nonlinear realm}

Differential changes of the extensive properties in the $K$th subspace are written as
\begin{eqnarray}
&&dE^K = \sum_i d(\epsilon_j^Kp_j^K)\\
&&dS^K = \sum_i\frac{d}{dt}(-p_j^K\ln \frac{p_j^K}{n_j^K})=\sum_i(-\ln \frac{p_j^K}{n_j^K}-1)\frac{dp_j^K}{dt}\quad
\end{eqnarray}
where $E^K$ and $S^K\equiv p^K\langle \tilde{s}\rangle^K$ are the energy and entropy in the $K$th subspace, respectively.

When a system is in a $M$th-order hypoequilibrium state and undergoes a pure relaxation process, a relation for property evolution in one subspace is acquired by using relation of Eq. (\ref{HES}), namely,
\begin{equation}
\frac{dS^K}{dt}=\sum(\epsilon_j^K\beta^K+\alpha^K-1)\frac{dp_j^K}{dt}
=\beta^K\frac{dE^K}{dt}+(\alpha^K-1)\frac{dp^K}{dt}
\end{equation}
The proof of this relation for one subspace applies to any differential changes (not only the time derivative). Thus, a generalization of the Gibbs relation to the $K$th subspace of a system in nonequilibrium, is expressed by
\begin{equation}
dS^K = \beta^KdE^K+(\alpha^K-1)dp^K
\end{equation}
Thus, from the Gibbs relation at stable equilibrium written as
\begin{equation}
dS = \frac{1}{T}dE-\frac{\mu}{T}dN
\end{equation}
the physical meaning of $\beta^K$ and $\alpha^K$ is shown to be
\begin{eqnarray}
&&\beta^K=\left(\frac{\partial S^K}{\partial E^K}\right)_{p^K}=\frac{1}{T^k}\\ &&\alpha^K-1=\left(\frac{\partial S^K}{\partial p^K}\right)_{E^K}=-\frac{\mu^K}{T^k},\,\mu^K=\left(\frac{\partial E^K}{\partial p^K}\right)_{S^K}
\end{eqnarray}
where $T^K$ is subspace temperature and $\mu^K$ is subspace chemical potential with respect to subspace probability $p^K$. The total differential entropy change for the system, which for a pure nonequilibrium relaxation process corresponds to the entropy generation, is
\begin{eqnarray}\label{Casimer}
dS &=& \sum_K dS^K= \sum_K \beta^KdE^K+\sum_K (\alpha^K-1)dp^K\nonumber\\
&=&\sum_K (\beta^K-\beta)dE^K+\sum_K (\alpha^K-\alpha)dp^K
\end{eqnarray}
where both energy ($\sum E^K=0$) and probability ($\sum p^K=0$) conservations have been applied. The Casimer condition holds and $J_{E}^K=dE^K/dt$ and $J_{p}^K=dp^K/dt$ are defined to be the internal fluxes of energy and probability inside the system, while $X_{p}^K=\beta^K-\beta$ and $X_{E}^K=\alpha^K-\alpha$ are the conjugate forces. The result is
\begin{eqnarray}
\sigma(\mathbf{J},\mathbf{X})=\frac{dS}{dt} = \sum_K X_{E}^KJ_{E}^K+\sum_KX_{p}^KJ_{p}^K\label{sigma}
\end{eqnarray}
The Onsager relations are then acquired from Eqs. (\ref{PE}) and (\ref{EE}) in the form of $\mathbf{J}=\mathbf{\Lambda} \mathbf{X}$, where $\mathbf{\Lambda}$ is a symmetric and positive definite operator. Thus, 
\begin{eqnarray}
J_p^K&=&\frac{1}{\tau}p^KX_{p}^K+\frac{1}{\tau}E^KX_{E}^K\label{Jp}\\
J_E^K&=&\frac{1}{\tau}E^KX_{p}^K+\frac{1}{\tau}\langle e^2\rangle^KX_{E}^K\label{JE}
\end{eqnarray}
The quadratic dissipation potential using force representation \cite{Gyarmati1970,Grmela2015} is written as
\begin{eqnarray}\label{Xi}
&&\Xi(\mathbf{X},\mathbf{X})=\frac{1}{2}\langle \mathbf{X},\Lambda \mathbf{X}\rangle=\frac{1}{2\tau}\sum_K[p^K(\alpha^K-\alpha)^2\nonumber\\&&+2E^K(\alpha^K-\alpha)(\beta^K-\beta)+\langle e^2\rangle^K(\beta^K-\beta)^2]
\end{eqnarray}
Using force representation, the variational principle is given by
\begin{eqnarray}
\delta[\sigma(\mathbf{J},\mathbf{X})-\Xi(\mathbf{X},\mathbf{X})]_\mathbf{J}=0,\, \mathbf{J}=const,\,\delta\mathbf{J}= 0,\,\delta\mathbf{X}\neq 0\nonumber\\
\end{eqnarray}
where the $\sigma(\mathbf{J},\mathbf{X})$ and $\Xi(\mathbf{X},\mathbf{X})$ are given by Eqs. (\ref{sigma}) and (\ref{Xi}). Furthermore, even though the following  constraints apply to the fluxes:
\begin{equation}
\sum_KJ_p^K=0,\,\sum_KJ_E^K=0
\end{equation}

The reciprocity seen in Eqs. (\ref{Jp}) and (\ref{JE}) is completely consistent with the Onsager theory since according to Gyarmati \cite{Gyarmati1970}, \textit{the validity of Onsager's reciprocal relations is not influenced by a linear homogeneous dependence valid amongst the fluxes}. Thus, the physical interpretation given here is fully consistent with other investigations \cite{LivonSpakovsky2015b} and does not require a reformulation in terms of independent fluxes even though this could be done. In addition, it is from the gradient dynamics of the nonequilibrium relaxation process that the entropy generation, the Onsager relations, and the quadratic dissipation potential of a local, isolated system in nonequilibrium have been derived using the geometric principle of SEA as well as the concepts of hypoequilibrium state and nonequilibrium intensive properties. Alternatively, the variational principle of SEA in system state space can be used to arrive at these relations as is done in \cite{Beretta2009} using the language of quantum mechanics. Of course, these relations also correspond to the variational principle in the space spanned by conjugate forces and fluxes \cite{Gyarmati1970}. 

\section{Theory: Heat diffusion in a nonequilibrium system}

A local, isolated system in nonequilibrium with heat diffusion only is considered in this section. This requires a model with a different set of constraints (i.e., the probability redistribution is only allowed in each subspace) then when only heat diffusion is considered. The entropy change of the system and subspaces due to heat diffusion for non-quasi-equilibrium process is given and the relationship between the SEAQT equation of motion and the phenomenological diffusion equation is presented.

\subsection{Equation of motion for heat diffusion}

Different from previous forms of the equation of motion, that for pure heat diffusion yields to a different set of conservation equations. If the system is separated into $M$ subspaces with energy flow but no probability flow across the subspaces, there are $M+1$ conservation laws. System probability conservation is replaced by that for $M$ individual subspaces. In Appendix A, it is proven that the concept of hypoequilibrium state and nonequilibrium temperature are also well defined under these new constraints given by
\begin{eqnarray}
I^K & = & \sum_{i} (x_i^K)^2= p^K,\,K=1,2,...M\\
E & = & \sum_{i} \epsilon_i x_{i}^2=\text{constant}
\end{eqnarray}
For simplicity, a second-order hypoequilibrium state is studied first. The system is separated into $2$ subspaces (subspace $a$ and subspace $b$) so that the equation of motion takes the form
\begin{equation}\label{HeatEM}
\frac{dp_j^a}{dt}=\frac{1}{\tau}\frac{\left|\begin{array}{cccc}
p_j^as_j^a & p_j^a & 0 & \epsilon_j^ap_j^a\\
\langle s\rangle^a & p^a & 0 & \langle e\rangle^a \\
\langle s\rangle^b & 0 & p^b & \langle e\rangle^b \\
\langle es\rangle & \langle e\rangle^a & \langle e\rangle^b & \langle e^2\rangle 
\end{array}\right|}{\left|\begin{array}{ccc}
p^a & 0 & \langle e\rangle^a \\
0 & p^b & \langle e\rangle^b \\
\langle e\rangle^a & \langle e\rangle^b & \langle e^2\rangle 
\end{array}\right|}
\end{equation}
where the contribution of each subspace to the total property is defined by
\begin{eqnarray}
\langle s\rangle = \langle s\rangle^a+\langle s\rangle^b,\,\langle s\rangle^a=\sum p_i^a s_i^a,\,\langle s\rangle^b=\sum p_i^b s_i^b\\
\langle e\rangle = \langle e\rangle^a+\langle s\rangle^b,\,\langle e\rangle^a=\sum p_i^a \epsilon_i^a,\,\langle e\rangle^b=\sum p_i^b \epsilon_i^b
\end{eqnarray}
where
\begin{equation}\label{s_j^a}
s_j^{a(b)} = -\ln \frac{p_j^{a(b)}}{n_j^{a(b)}}=\tilde{s}_j^{a(b)}-\ln p^{a(b)},\,\tilde{s}_j^{a(b)}\equiv-\ln \frac{\tilde{p}_j^{a(b)}}{n_j^{a(b)}}
\end{equation}
By defining
\begin{eqnarray}
B_1=\left|\begin{array}{ccc}
p^a & 0 & \langle e\rangle^a \\
0 & p^b & \langle e\rangle^b \\
\langle e\rangle^a & \langle e\rangle^b &\langle e^2\rangle 
\end{array}\right|,
B_2^a=\left|\begin{array}{ccc}
\langle s\rangle^a & 0 &\langle e\rangle^a \\
\langle s\rangle^b & p^b & \langle e\rangle^b \\
\langle es\rangle & \langle e\rangle^b & \langle e^2\rangle 
\end{array}\right|\nonumber\\
B_2^b=\left|\begin{array}{ccc}
\langle s\rangle^b & 0 &\langle e\rangle^b \\
\langle s\rangle^a & p^a & \langle e\rangle^a \\
\langle es\rangle & \langle e\rangle^a & \langle e^2\rangle 
\end{array}\right|,
B_3=\left|\begin{array}{ccc}
\langle s\rangle^a & p^a & 0 \\
\langle s\rangle^b & 0 & p^b \\
\langle es\rangle & \langle e\rangle^a & \langle e\rangle^b 
\end{array}\right|
\end{eqnarray}
Eq. (\ref{HeatEM}) can be simplified to
\begin{equation} \label{EQMH}
\frac{dp_j^a}{dt}=\frac{1}{\tau}(p_j^as_j^a-p_j^a\frac{B_2^a}{B_1}-\epsilon_j^ap_j^a\frac{B_3}{B_1})
\end{equation}
Moreover, the equation of motion for the probability distribution in one subspace can also be written in terms of the normalized probability by dividing both sides of Eq. (\ref{EQMH}) by $p^a$ so that
\begin{equation}\label{HeatEQ}
\frac{d\tilde{p}_j^a}{dt}=\frac{1}{\tau}(\tilde{p}_j^as_j^a-\tilde{p}_j^a\frac{B_2^a}{B_1}-\epsilon_j^a\tilde{p}_j^a\frac{B_3}{B_1})
\end{equation}
Furthermore, if the system is in a second-order hypoequilibrium initially so that each subspace has a canonical distribution, Eq. (\ref{HeatEQ}) can be simplified further to arrive at the form 
\begin{equation}
\frac{d\tilde{p}_j^a}{dt}=\frac{1}{\tau}\tilde{p}_j^a[(\tilde{s}_j^a-\langle \tilde{s}\rangle^a)-\beta(\epsilon_j^a-\langle \tilde{e}\rangle^a)]
\end{equation}
where $\tilde{s}_j^a$ is defined by Eq. (\ref{s_j^a}) and $\langle \tilde{s}\rangle^a$ and $\langle \tilde{e}\rangle^a$ are defined by Eqs. (\ref{eK}) and (\ref{sK}). The parameter $\beta$ is given
\begin{eqnarray}\label{beta}
\beta&&\equiv\frac{B_3}{B_1}=\frac{p^a\tilde{A}_1^a\beta^a+p^b\tilde{A}_1^b\beta^b}{p^a\tilde{A}_1^a+p^b\tilde{A}_1^b}\\
B_1&&=p^ap^b(p^a\tilde{A}_1^a+p^b\tilde{A}_1^b)\\
B_3&&=p^ap^b(p^a\beta^a\tilde{A}_1^a+p^b\beta^b\tilde{A}_1^b)
\end{eqnarray}
is a weighted average of the inverse temperatures of the subsystems relative to the mole fractions and the energy fluctuation (or nondimensional specific heat at constant volume) of the subspaces written as
\begin{eqnarray}
&&\tilde{A}_1^{a(b)} = \langle\tilde{e^2}\rangle^{a(b)} - (\langle \tilde{e}\rangle^{a(b)})^2=-\frac{\partial\langle\tilde{e}\rangle^{a(b)}}{\partial\beta^{a(b)}} = \frac{C_V^{a(b)}}{(\beta^{a(b)})^2}\quad\\
&&C_V^{a(b)}\equiv\frac{1}{k_b}\frac{\partial\langle\tilde{e}\rangle^{a(b)}}{\partial T^{a(b)}}
\end{eqnarray}
For the more general case of a $M$th-order hypoequilibrium state and the system separated into $M$ subspaces, equation (\ref{HeatEQ}) remains the same but with
\begin{equation}
\beta\equiv\frac{B_3}{B_1}=\frac{\sum_K^M p^K\tilde{A}_1^K\beta^K}{\sum_K^M p^K\tilde{A}_1^K}
\end{equation}

A given interaction type (e.g., heat diffusion) results in a given relaxation time $\tau$ (see Sec. II.B), while the ratio $B_3/B_1$ provides an average temperature based on subspace mole fractions and energy fluctuations. At stable equilibrium, $\beta=\beta^{eq}$. Now, if one subspace $R$ is attached to a reservoir, the evolution of the other subspaces behave according to the equation of motion, Eq. (\ref{HeatEQ}), with $\beta^R$ constant. For example, only part of the energy eigenlevels can absorb energy from the environment. Mathematically, if subspace $R$ yields to one of two conditions,
\begin{equation}
\forall K\neq R, C_V^R\gg C_V^K, \quad p^R\gg p^K
\end{equation}
the relation $\beta = \beta^R$ holds and subspace $K(\neq R)$ yields to the equation of motion.
\begin{equation}
\frac{d\tilde{p}_j^K}{dt}=\frac{1}{\tau}\tilde{p}_j^K[(-\ln \frac{\tilde{p}_j^K}{n_j^K}-\langle \tilde{s}\rangle^K)-\beta^R(\epsilon_j^K-\langle \tilde{e}\rangle^K)]
\end{equation}
Note that in this equation, the only parameter related to subspace $R$ is the reservoir temperature $\beta^R$. The energy eigenstructure of subspace $R$ plays no role!

\subsection{Property of heat diffusion: non-quasi-equilibrium process and second law of thermodynamics}

Based on Eq. (\ref{EQMH}), the total entropy and energy evolution in one subspace can be determined via
\begin{eqnarray}
\frac{dS^K}{dt}&=&\frac{d\langle s\rangle^K}{dt}=p^K\frac{d\langle \tilde{s}\rangle^K}{dt}=\frac{1}{\tau}p^K(\beta^K-\beta)\beta^K\tilde{A}_1^K\quad\label{73}\\
\frac{dE^K}{dt}&=&\frac{d\langle e\rangle^K}{dt}=p^K\frac{d\langle \tilde{e}\rangle^K}{dt}=\frac{1}{\tau}p^K(\beta^K-\beta)\tilde{A}_1^K\label{74}
\end{eqnarray}
Dividing Eq. (\ref{73}) by (\ref{74}) yields
\begin{equation}
\frac{dS^K}{dE^K}=\frac{dS^K}{dt}/\frac{dE^K}{dt}=\beta^K
\end{equation}
The equation is a generalized form of the differential entropy transfer due to heat diffusion using the nonequilibrium temperature for each subspace, i.e.,
\begin{equation}\label{q}
dS^K=\beta^{K}dE^K=\frac{\delta Q^K}{T}
\end{equation}
Moreover, Eq. (\ref{q}) can applied to all kinds of thermodynamic processes, and is not limited to quasi-equilibrium processes. This argument comes from the universal definition of nonequilibrium temperature provided in this paper.

\section{Model: Composite system in a nonequilibrium state}

Interacting systems can form a composite nonequilibrium system with the interaction resulting in the nonequilibrium relaxation process for the composite. Using the SEA equation of motion, the state evolution of the composite system can be determined. Proper division of this composite allows the subspaces to be viewed as the interacting subsystems within the composite. The behavior of each subsystem, thus, can be studied by an analysis of the state evolution of each subspace. In particular, if two individual subsystems involved in an interaction both have canonical state distributions, and each subsystem’s energy eigenvectors span one subspace, the composite system is in a second–order hypoequilibrium state. On the other hand, if each individual subsystem's state cannot be described by a canonical distribution, a higher-order hypoequilibrium state will be required.

In the following section, (Sec. V.A), the SEAQT framework using single particle energy eigenlevels is applied to the study of a simple system. Section V.B then explains the process of subspace division followed in Sec. V.C by a comparison with the phenomenological equations for mass and heat diffusion. In Sec. V.D, the physical details of the system used are described. Finally, in Sec. V.E, the coupling of mass and heat diffusion is modeled and discussed.

\subsection{Multi-particle classical simple system}
Theoretically and in general, the SEAQT equation of motion is applicable to multi-particle systems provided energy eigenstructure of the system is known \cite{LivonSpakovsky2015b}. However, for a multi-particle classical simple system, the energy eigenstructure of a single particle and its associated equation of motion can be used to study the system, since all of the particles (or particle groups) have the same energy eigenlevels $\{\epsilon_i, i=1,2,3,...\}$ and degeneracy $\{n_i, i=1,2,3,...\}$. Thus, the system state can be represented by the particle number (or particle group number) at each energy eigenlevel $\{m_i, i=1,2,3,...\}$. The mole fraction of particles at the $i$th energy eigenlevel is given by $y_i=m_i/\sum m_i$. The extensive property constraints of the system are then
\begin{eqnarray}
	M &=&\sum_{i} m_i=\text{constant}\\
	E &=&\sum_{i} \epsilon_i m_i=\text{constant}\\
	S &=&\sum_{i} -m_i \ln\frac{y_i}{n_i}
\end{eqnarray}
Dividing the constraints by the total particle number $\sum m_i$, the system state can be represented by the mole fractions $\{y_i, i=1,2,3,...\}$, which is equivalent to the single-particle probability distribution $\{p_i, i=1,2,3,...\}$. For an isolated system the constraints become
\begin{eqnarray}
	I &=&\sum_{i} y_i=\sum_{i} p_i=\text{constant}\\
	\langle e \rangle &=& \frac{E}{M}=\sum_{i} \epsilon_i y_i=\sum_{i} \epsilon_i p_i=\text{constant}\\
	\langle s \rangle &=& \frac{S}{M}=\sum_{i} -y_i \ln\frac{y_i}{n_i}=\sum_{i} -p_i \ln\frac{p_i}{n_i}
\end{eqnarray}
$\langle\dots\rangle$ represents an average specific property. The constraints and equation of motion reduce to the single-particle case (Eqs. (\ref{5}) to (\ref{7}) and (\ref{SEAQTEM})) when $y_i=p_i, i=1,2,3,...$.

In general, a system’s energy eigenstructure and extensive properties are a function of the total particle number of each constituent. This is also true of its specific properties in the presence of a mass interaction or chemical reaction if the system is partitioned and not simple \cite{Gyftopoulos2005}, since partitioning influences each partition's energy eigenstructure. Thus, for mass diffusion, the framework outlined here requires an invariant eigenstructure and as a consequence the simple system assumption. With this assumption, particle number no longer influences the specific properties. This same assumption, however, is not required in the case of heat diffusion since the total number of particles for each system partition (i.e., subsystem) does not change.

\subsection{Interacting systems}
It is assumed that a group of observable operators $\hat{F}$ commuting with the Hamiltonian operator $\hat{H}$ exists. The degenerate energy eigenlevels of the system can be distinguished by eigenvalues of the observations of $\hat{F}$, which have values $F_1, F_2,\cdots,F_M$, so that the system energy eigenlevels can be separated into $M$ sets $\{\epsilon_i^K,i=1,2,3,...\}$ with degeneracy $\{n_i^K,i=1,2,3...\}$, where $K=1,\cdots,M$. In each set, every energy eigenlevel represents an eigenstate common to both $\hat{F}$ and $\hat{H}$ with the same eigenvalue of $\hat{F}$. Eigenstates in each of the sets can be spanned into a subspace of the system state space and can be designed as a subsystem. 

\begin{figure}
	\centering
	\includegraphics[width=3.3in]{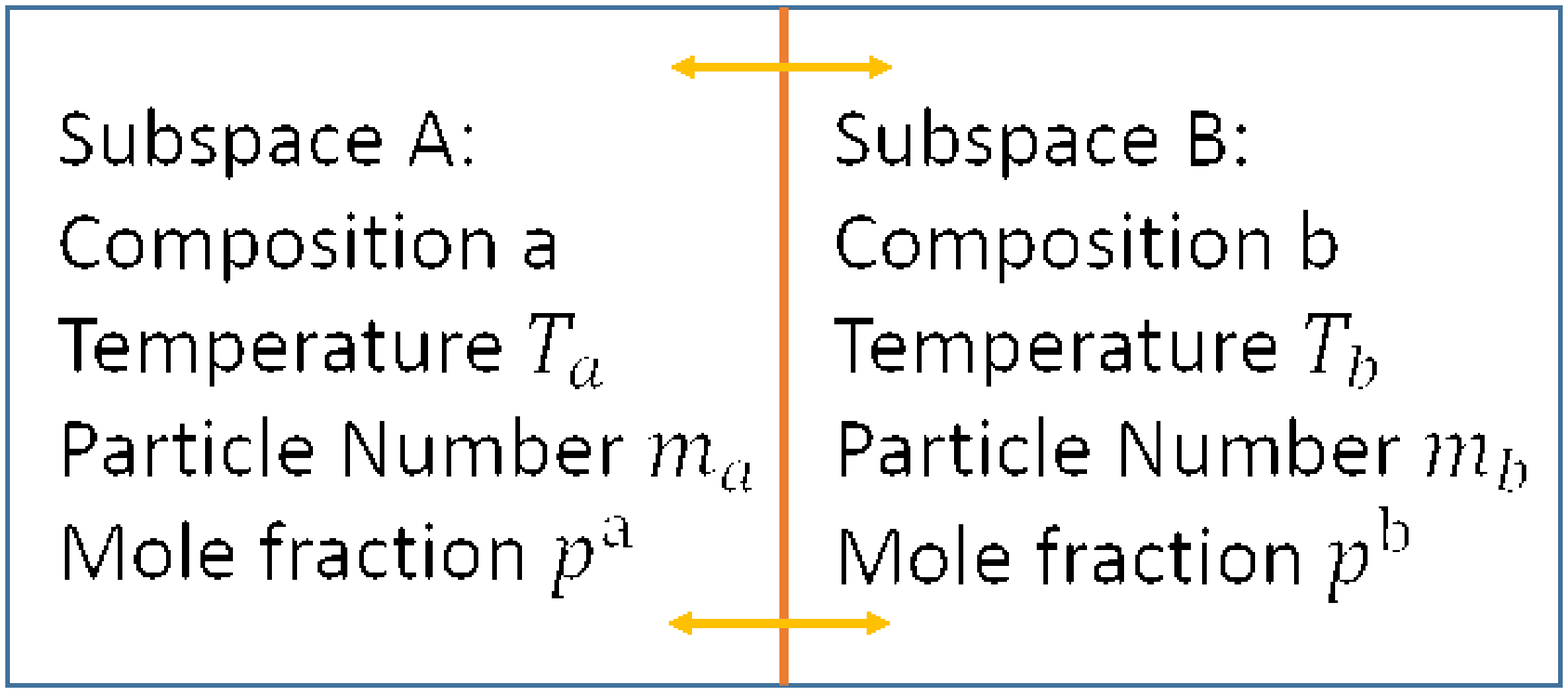}
	\caption[Partition in simple system]{For mass diffusion, mass flow and energy flow are both allowed across the partition. For heat diffusion, only energy flow is allowed.}
\end{figure}

Practically, by choosing the observable operator, the system can be view as an composite system whose subsystems can be properly arranged to study specific phenomena. For the kinetics of a chemical reaction, $F$ can be chosen to be an observable operator of species, whose eigenvalues are $``Reactant"$ and $``Product"$. In heat and mass diffusion, the observable operator $F$ is chosen to be the relative location to a partition, with eigenvalue of $``left"$ (left of the partition) and $``right"$ (right of the partition). The partition allows mass and heat diffusion (see Fig. 1).%Alternatively, two position values is used as a rough approximation of continuous eigenvalues of position operator. This subspaces division needs the position operator and Hamiltonian operator commute. 
To be more precise, it is assumed that the de Brogli wavelength $\lambda$ is much smaller than the distance $\Delta L$ between the center of the subsystem on the $``left"$ and that on the $``right"$. This wavelength represents the classical limit, i.e.,
\begin{equation}
	p_v=\sqrt{2mk_bT},\quad\lambda=\frac{h}{p_v}=\frac{h}{\sqrt{2mk_bT}},\quad \Delta L\gg\lambda
\end{equation}
where $p_v$ is the expectation value of the particle momentum for a system at temperature $T$. 

\subsection{Phenomenological transport equation}
With the assumption that the two subspaces of the system are two subsystems at two positions, the phenomenological transport equations of mass and heat diffusion can be derived.

\subsubsection{Mass diffusion}
Using Eq. (\ref{28}), the equations of motion for two subspaces are written as
\begin{eqnarray}
\frac{d\ln p^a}{dt}=-\frac{1}{\tau}\ln p^a+\frac{1}{\tau}(\langle \tilde{s}\rangle^a-\frac{A_2}{A_1}+\langle \tilde{e}\rangle^a\frac{A_3}{A_1})\label{84}\\
\frac{d\ln p^b}{dt}=-\frac{1}{\tau}\ln p^b+\frac{1}{\tau}(\langle \tilde{s}\rangle^b-\frac{A_2}{A_1}+\langle \tilde{e}\rangle^b\frac{A_3}{A_1})\label{85}
\end{eqnarray}
When the two subspaces of the system have the same eigenstructure and temperature, subtracting Eq. (\ref{85}) from Eq. (\ref{84}) yields
\begin{equation}\label{86}
\frac{d}{dt}(\ln p^a-\ln p^b) = -\frac{1}{\tau}(\ln p^a-\ln p^b)
\end{equation}
Substituting the subspace probability (or mole fraction), Eq. (\ref{87}) into Eq. (\ref{86}) results in
\begin{eqnarray}
p^a = \frac{n^a}{n^a+n^b} \quad p^b = \frac{n^b}{n^a+n^b}\label{87}\\
\frac{d}{dt}(\ln \frac{n^a}{n^b}) = -\frac{1}{\tau}(\ln \frac{n^a}{n^b})\label{88}
\end{eqnarray}
If the global mass distribution is continuous, and position A and position B are close enough,
\begin{equation}
n^a = n^b+\Delta n
\end{equation}
which transforms Eq. (\ref{88}) into
\begin{eqnarray}
&&\frac{d}{dt}(\frac{\Delta n}{n^b}) = -\frac{1}{\tau}\frac{\Delta n}{n^b}\label{90}\\
&&J^{a(b)}\equiv\frac{1}{2A}\frac{d}{dt}(n^a-n^b) = -\frac{n^a-n^b}{2\tau A}=-\frac{\delta x}{2\tau A}\frac{dn}{dx}\label{91}
\end{eqnarray}
where the approximation $\ln(1+x)\simeq x$ for small $x$ has been used and higher-order terms dropped to arrive at Eq. (\ref{90}). In Eq. (\ref{91}), $\delta x$ is the distance between two positions, $A$ is the cross-sectional area of the interacting surface, and $J^{a(b)}$ is the flow of particle number equal to $(dn^a/dt)/A$ or $(-dn^b/dt)/A$ (hence the division by $2$ in Eq. (\ref{91})). Eq. (\ref{91}) recovers Fick's law with the diffusion coefficient (diffusivity) given by $D = \frac{\delta x}{2\tau A}$. The specific form of $D$ is directly related to the form of $\tau$, which contains the detailed mechanical information. The phenomenological linear equation can be derived without the form of $\tau$, which is a pure thermodynamic feature or pattern of the nonequilibrium relaxation process. In addition to results such as these for the near-equilibrium realm, thermodynamic features or patterns in the far-from-equilibrium realm can also be studied using Eqs. (\ref{PE}) and (\ref{EE}) provided the initial state is a hypoequilibrium state. For the case when it is not, Eq. (\ref{SEAQTEM}) can directly be used as is done, for example, in \cite{LivonSpakovsky2015}.

\subsubsection{Heat diffusion}

For system in which the only interaction is that of   heat diffusion, Eq. (\ref{q}) captures the energy flow between two subsystems, i.e.,
\begin{eqnarray}\label{92}
\dot{Q}&=&J^{a(b)}_E=-\kappa'\delta xA\frac{dT}{dx}=\frac{1}{\tau}p^a(\beta^a-\beta)\tilde{A}_1^a\nonumber\\
&=&\frac{1}{\tau}\frac{p^ap^b\tilde{A}_1^a\tilde{A}_1^b}{p^a\tilde{A}_1^a+p^b\tilde{A}_1^b}(\beta^a-\beta^b)
\end{eqnarray}
where $\dot{Q}$ is the rate of energy transferred, $J^{a(b)}_E$ is the heat flux, $A$ is the cross-sectional area of the interacting surface, $T^a$ and $T^b$ are temperature of the two subsystems, and Eq. (\ref{beta}) has been substituted for $\beta$. Eq. (\ref{92}) recovers Fourier’s law of heat diffusion (conduction). The thermal conductivity per unit length $\kappa'$ and thermal conductivity $\kappa$ are expressed as: 
\begin{equation}
\kappa'=\frac{1}{\tau}\frac{p^ap^b\beta^a\beta^b\tilde{A}_1^a\tilde{A}_1^b}{p^a\tilde{A}_1^a+p^b\tilde{A}_1^b}\frac{1}{A}, \quad \kappa=\kappa'\delta_x
\end{equation}
In the near-equilibrium region with the same constituent in the two subsystems,
\begin{equation}
p^a=p^b=\frac{1}{2}, \quad C_V^a = C_V^b=C_V
\end{equation}
and the thermal conductivity per unit length and the thermal conductivity are expressed in terms of the energy fluctuation (or nondimensional specific heat at constant volume) of the subspaces, i.e.,
\begin{equation}
\kappa'=\frac{1}{2\tau}C_V\frac{1}{A}, \quad \kappa=\frac{1}{2\tau}\frac{\delta_x}{A}C_V
\end{equation}
The above formulation is applicable for any kind of interaction resulting in a flow of energy only. Furthermore, if the heat and mass diffusion are affected via the same kind of micro-mechanical interactions such as the collision of particles, it can be assumed that the same $\tau$ is applicable when the system is in the near-equilibrium region near to the same stable equilibrium point. In this case, $\kappa=C_VD$. This last result is the same as that found from classical transport theory and is a direct consequence of the thermodynamic features of the system minus any direct knowledge of the details of the micro-mechanical interactions taking place.

\subsection{Mass and heat diffusion of hydrogen}
To model the mass and heat diffusion for a specific case, a composite system of hydrogen is set up with two subspaces corresponding to subsystems on two sides of a partition. The energy eigenlevels of the two subsystems together form the energy eigenlevels for the composite system as a whole. Denoting the state space of the subsystem on the $``left"$ by $\mathcal{H}^{a}$ and that on the $``right"$ by $\mathcal{H}^{b}$, the composite system state space $\mathcal{H}$ takes the form

\begin{equation}
\mathcal{H}=\mathcal{H}^{a}\oplus\mathcal{H}^{b}
\end{equation}
The available energy eigenvalues for one subspace ($``left"$ or $``right"$) are constructed from the energy eigenvalues of each degree of freedom for translation and rotation, i.e., from
\begin{eqnarray}
\epsilon^{a(b)}=\epsilon_{t,H_2}+\epsilon_{r,H_2}
\end{eqnarray}
The translational energy eigenvalue $\epsilon_t$ uses the form of the infinite potential well, while the rotational energy eigenvalue $\epsilon_r$ uses the form of the rigid motor. These are expressed as follow:
\begin{eqnarray}
&&\epsilon_t(n_x,n_y,n_z)=\frac{\hbar^2}{8m}\left(\frac{n_x^2}{L_x^2}+\frac{n_y^2}{L_y^2}+\frac{n_z^2}{L_z^2}\right)\\
&&\epsilon_r(j,m)=\frac{j(j+1)\hbar^2}{2I}
\end{eqnarray}
where $n_x$, $n_y$, and $n_z$ are the quantum numbers for the translational degrees of freedom, $j$ and $m$ are the quantum numbers for the rotational degrees of freedom, $I$ is the moment of inertia and $L_x$, $L_y$, and $L_z$ are chosen based on mean-free-path of the particles in each subsystems (e.g., the dimension of subsystem is used for an ideal gas). The vibrational energy are not included since the temperature in the study is below its characteristic temperature. For more discuss on the vibrational energy, reader is refereed to \cite{LivonSpakovsky2015}. The disassociated energy is not included by selection of the proper energy reference. Each combination of quantum numbers and position corresponds to one energy eigenlevel in the subspaces (or subsystems). The composite system energy eigenlevels are formed by all the available energy eigenlevels of the $``left"$ and the $``right"$.

A second-order hypoequilibrium state with the subspace division of $``left"$ and $``right"$ is chosen to be the initial condition, which means that two subsystems are in local equilibrium states. The non-quasi-equilibrium process of mass and heat diffusion is studied using Eqs. (\ref{EQM}) and (\ref{EQMH}). For the case when the subsystems are not in states of local equilibrium, the two subspaces can be divided further. For example, if the $``left"$ subsystem is a $M$th-order hypoequilibrium state, this subspace, subspace $a$, can be divided into $M$ subspaces according to the initial condition. However, the evolution of each subspace, regardless of whether or not the subsystem is in a state of local equilibrium, yields to the same form of the equations of motion, Eqs. (\ref{EQM}) and (\ref{EQMH}). For the case considered here, the initial condition is given by (\ref{HEPEi}), i.e., by

\begin{eqnarray}
p_i^{a(b)}(t=0)&=&\frac{p^{a(b)}n_i^{a(b)}}{Z^{a(b)} (\beta^{a(b)})} e^{-\beta^{a(b)}\epsilon_i^{a(b)}}
\end{eqnarray}
The time evolution is acquired by solving Eqs. (\ref{EQM}) and (\ref{EQMH}). For a more general initial condition, such as that for an infinite-order hypoequilibrium state, the equation of motion can be solved using the density of states method developed in \cite{LivonSpakovsky2015}.

The specific properties of the individual subsystems at a given temperature and volume are given
\begin{eqnarray}\label{101}
Z^{a(b)}(\beta^{a(b)},V)&=&Z^tZ^r=V(\frac{m}{2\pi\hbar^2\beta^{a(b)}})^\frac{3}{2}\frac{2I}{\beta^{a(b)}\hbar^2}\nonumber\\
&=&C_ZV(\beta^{a(b)})^{-5/2}\\
\langle\tilde{e}\rangle^{a(b)}(\beta^{a(b)})&=&\frac{5}{2}k_bT^{a(b)}=\frac{5}{2\beta^{a(b)}}=\frac{C_V}{\beta^{a(b)}}\label{102}\\
\langle\tilde{s}\rangle^{a(b)}(\beta^{a(b)},V)&=&\beta^{a(b)}\langle\tilde{e}\rangle^{a(b)}+\ln Z^{a(b)}\nonumber\\
&=&-\frac{5}{2}\ln\beta^{a(b)}+\ln V+C_s\nonumber\\
&=&-C_V\ln\beta^{a(b)}+\ln V+C_s\label{103}
\end{eqnarray}
where $C_Z$ and $C_s$ are constants determined from Eqs. (\ref{101}) to (\ref{103}).

\subsection{Mass and heat diffusion coupling}
For mass diffusion with no temperature difference, as shown in Eq. (\ref{bE}), the temperature difference remains zero, and only the particle number difference changes and Eq. (\ref{PE}) reverts to the linear transport equation in the near-equilibrium realm. The more complex case occurs when mass diffusion takes place in the presence of a temperature difference is of interest in which case coupling effects may be present. 

The mass flow between two subsystems is determined by subtracting the probability evolution of one [Eq. (\ref{PE})] from the other with the result that.
\begin{eqnarray}\label{104}
2J_{p}^{b\to a}&&=\frac{dp^a}{dt}-\frac{dp^a}{dt}=\frac{1}{\tau}p^{a}(\alpha^a-\alpha)+\frac{1}{\tau}p^{a} \langle \tilde{e}\rangle^{a}(\beta^a-\beta)\nonumber\\
&&-\frac{1}{\tau}p^{b}(\alpha^b-\alpha)-\frac{1}{\tau}p^{b} \langle \tilde{e}\rangle^{b}(\beta^b-\beta)
\end{eqnarray}
The coupling effects come from the differences in both of the nonequilibrium intensive properties $\alpha$ and $\beta$. To study the effect of temperature on the probability (mass) flow, the two subsystems start from the same initial probability ($p^a=p^b=p^{eq}=0.5$) but different temperatures. For the case of a perfect gas (i.e., ideal gas with constant specific heats), the final stable equilibrium temperature $T^{eq}$ is an average of the initial temperatures of the two subsystems, namely,
\begin{eqnarray}
&&T^{eq}=(T^a+T^b)/2,\,\xi\equiv\Delta T/T^{eq}\nonumber\\
&&T^a=T^{eq}+\Delta T,\, T^b=T^{eq}-\Delta T,\, 
\end{eqnarray}
Thus, Eq. (\ref{104}) can be simplified to
\begin{eqnarray}
J_{p}^{b\to a}&&=\frac{1}{2\tau}p^{eq}[(\langle \tilde{s}\rangle^{a}-\langle \tilde{s}\rangle^{b})-\beta(\langle \tilde{e}\rangle^{a}-\langle \tilde{e}\rangle^{b})]
\end{eqnarray}
where $\langle \tilde{s}\rangle^{a(b)}$ and $\langle \tilde{e}\rangle^{a(b)}$ are the specific entropy and energy of subsystem a(b) defined by Eqs. (\ref{eK}) and (\ref{sK}). Substituting Eqs. (\ref{102}) and (\ref{103}) yields
\begin{eqnarray}
J_{p}^{b\to a}&&=\frac{1}{2\tau}p^{eq}[C_V(\ln\frac{1}{\beta^a}-\ln\frac{1}{\beta^b})-\beta(\frac{C_V}{\beta^a}-\frac{C_V}{\beta^b})]\,\nonumber\\
&&=\frac{C_Vp^{eq}}{2\tau}[(\ln T^a-\ln T^b)-\frac{1}{T}(T_a-T_b)]
\end{eqnarray}
where $T=1/(k_b\beta)$ and the relation $T/T^{eq}=1+\xi^2+O(\xi^4)$ holds when $p^a=p^b$. The mass (probability) flux due to a temperature difference is then written as
\begin{eqnarray}\label{108}
J_{p}^{b\to a}&&=\frac{dp^a}{dt}\frac{C_Vp^{eq}}{\tau}[\frac{\Delta T}{T^{eq}} +\frac{\Delta T^3}{(T^{eq})^3}-\frac{\Delta T}{T}] \nonumber\\
&&=\frac{4}{3}\frac{C_Vp^{eq}}{\tau}(\frac{\Delta T}{T^{eq}})^3+O(\lambda^5)
\end{eqnarray}
and the temperature evolution, Eq. (\ref{bE}), reduces to
\begin{equation}\label{109}
\frac{dT^a}{dt}=-\frac{1}{\tau}\Delta T
\end{equation}
\begin{figure}
	\center
	\includegraphics[width=3.3in]{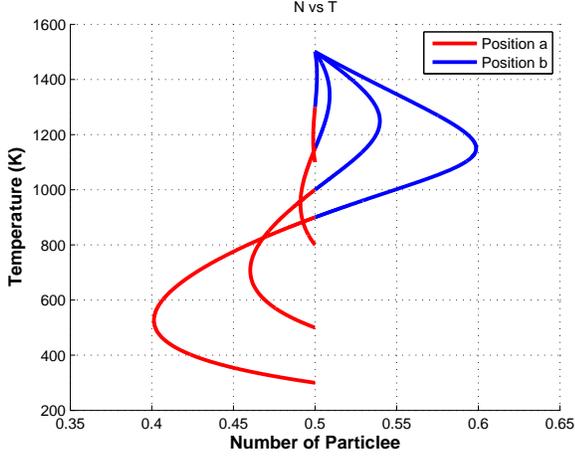}
	\caption[(color online) Thermodynamic trajectories on a temperature-particle number diagram with only temperature difference, Change figure]{Thermodynamic trajectories on a temperature-particle number diagram. The initial probabilities for the two subsystems are the same, while the initial temperature of subsystem $b$ is $1500K$, while that for subsystem $a$ is $300$ K, $500$ K, $800$ K and $1100$ K, respectively.}
\end{figure}

In the near-equilibrium region where only small temperature differences exist ($\xi\ll 1$), higher-order nonlinear temperature difference effects, which influence the probability (mass) flux, are negligible and can, thus, be ignored. For this case, the temperature evolution equation [Eq. (\ref{109})] and the probability (mass) evolution equation [Eq. (\ref{108})] due to a temperature difference are effectively decoupled. In far-from-equilibrium realm, however, higher-order temperature difference nonlinearities may be significant in which case coupling effects become important. In Fig. 2, the thermodynamic trajectories of three different cases for which the initial probabilities for the two subsystems are the same are plotted on a temperature-particle number diagram. For each case, the trajectory consists of two lines, one for each subsystem, with each point on each line representing an intermediate state for a given subsystem. The two subsystems start from opposite ends of the two colored lines and evolve towards the common end of the lines, which is the state of stable equilibrium for the system. As can be seen in the figure, when the temperature difference is small, the maximum of the concentration difference through the evolution approaches zero very quickly. Since the lowest order terms of Eq. (\ref{108}) and (\ref{109}) have different signs, the nonlinear effects of temperature drive the probability (mass) flux towards the higher temperature subsystem. This phenomena can be explained from an entropy generation standpoint. The higher temperature subsystem has a higher specific entropy so that the probability (mass) flux towards it results in entropy generation for the system. On the other hand, the temperature evolution is explained by the fact that the heat diffusion towards the lower temperature subsystem increases the specific entropy in the lower temperature subsystem, which in turn results in entropy generation for the system.

\begin{figure}
	\centering
	\includegraphics[width=3.3in]{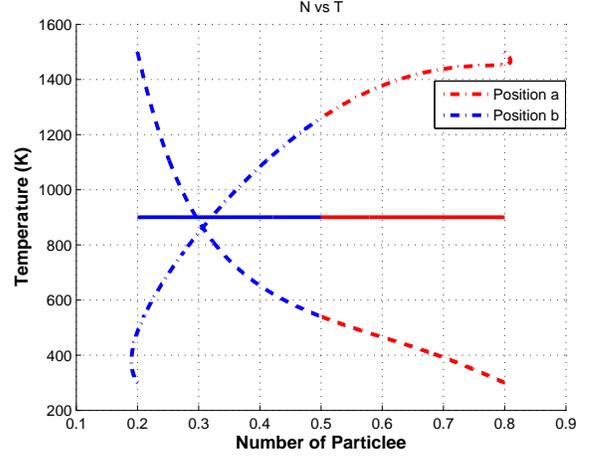}
	\caption[Thermodynamic trajectories on a temperature-particle number diagram with probability and temperature difference] {(color online)Thermodynamic trajectories on a temperature-particle number diagram for three cases. Case 1: $p^a=0.8$, $p^b=0.2$,$ T^a=1500$ K, $T^b=300$ K (dashed-dotted line); Case 2: $p^a=0.8$, $p^b=0.2$, $T^a=900$ K, $T^b=900$ K (solid line); and Case 3: $p^a=0.2$, $p^b=0.8$, $T^a=1500$ K, $T^b=300$ K (dashed line). The red lines are the trajectory of subsystem $a$, and the blue lines are the trajectory of subsystem $b$.}
\end{figure}
\begin{figure}
	\centering
	\includegraphics[width=3.3in]{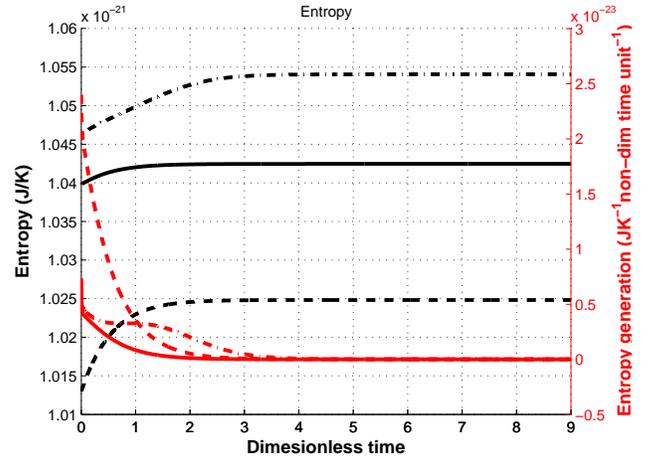}
	\caption[Entropy evolution and entropy generation with probability and temperature difference]{(color online) Entropy evolution and entropy generation in dimensionless time of the three cases of Fig. 4.3. Case 1: $p^a=0.8$, $p^b=0.2$,$ T^a=1500$ K, $T^b=300$ K (dashed-dotted line); Case 2: $p^a=0.8$, $p^b=0.2$, $T^a=900$ K, $T^b=900$ K (solid line); and Case 3: $p^a=0.2$, $p^b=0.8$, $T^a=1500$ K, $T^b=300$ K (dashed line). The black lines are the entropy evolutions relative to the vertical axis on the left, and the red lines are the entropy generation rates relative to the vertical axis on the right.}
\end{figure}

When probability (mass) and temperature differences exist at the same time, it is the combined effect (i.e., the coupling) from the probability and the temperature, which determines the probability flow, since the lower order terms of Eqs. (\ref{91}) and (\ref{108}) have opposite signs. In Fig. 3, the trajectory for Case 1 shows a competition effect between the probability and temperature, while the trajectory of Case 3 shows a cooperation effect. Both of these can be explained via the effects discussed relative to Fig. 2. in addition, as a validation, Fig. 4 provides the entropy generation rate for the three cases along with the entropy trajectories. All exhibit monotonic increases in the entropy over time. Comparing to the diffusion case (Case 2) without temperature difference, the case for which the temperature effect is competitive (Case 1) results in a greater variation in the entropy generation rate, while the case for which the temperature effect is cooperative results in a much steeper drop in the entropy generation rate.
\section{Conclusion}
This paper investigates the relaxation process of local, isolated systems in nonequilibrium using the SEAQT framework. The mass and heat diffusion inside the system, which are the mass and energy redistribution among the system subspaces (or equivalent, energy eigenlevels), are described by defining conjugate forces and conjugate fluxes using the concepts of hypoequilibrium state and nonequilibrium intensive properties. These thermodynamic features or patterns of the nonequilibrium relaxation process are used to generalize the Gibbs relation, the Clausius inequality, and the Onsager relations to the far-from-equilibrium realm and for quasi-nonequilibrium processes. The variational principle in the spaces spanned by conjugate forces and conjugate fluxes is derived from the variational principle in system state space (i.e., from the principle of steepest entropy ascent). As an application, the mass diffusion of a simple system consisting of hydrogen is studied. The study results in decoupled mass and energy transport equations and their associated phenomenological coefficients in the near-equilibrium realm. In far-from-equilibrium realm, the coupling phenomena and the nonlinear effects for mass and energy transport are derived.

From this investigation it is evident that the introduction of the concepts of hypoequilibrium state and nonequilibrium intensive properties into the SEAQT framework provides a novel and fundamental vantage point from which to describe nonequilibrium states and their evolution during a relaxation process. In addition to the study presented here, additional work using the density of states method developed in \cite{LivonSpakovsky2015}, has permitted the wide application of the SEAQT framework to the study of nonequilibrium, systems in which complex, coupled reaction diffusion pathways are modeled and compared with experiment \cite{LiASME2015b,LiJPS2015}. chemical reaction, mass and heat diffusion. As a complement to the present paper, \cite{LivonSpakovsky2015b} continues our study of the non-quasi-equilibrium process of two interacting systems and completes the discussion of Onsager type of investigation of the relaxation process with both fluxes inside a nonequilibrium system and those across different systems. All of these studies show SEAQT to be a powerful approach applicable to the study of nonequilibrium phenomena across all temporal and spatial scales.

\section*{Acknowledgment}
Funding for this research was provided by the US Office of Naval Research under Award No. N00014-11-1-0266.
\appendix	
\section{Hypoequilibrium for heat diffusion}
In this appendix, it is proven that for a system with heat diffusion only, if the initial state is given by Eq. (23), the system evolution is also given by Eq. (\ref{HEPE}), and the nonequilibrium temperature is well-defined. The proof follows the same process as in \cite{LivonSpakovsky2015} for a system with probability and energy conservations. To show this, Eq. (62) is reformulated such that
\begin{equation}
\frac{d}{dt}\ln\frac{p_j^a}{n_j^a}=\frac{1}{\tau}(-\ln\frac{p_j^a}{n_j^a}-\frac{B^a_2}{B_1}+\epsilon_j^a\frac{B_3}{B_1})
\end{equation}
where it is noted that $d(\ln n_j^a)/dt$ is zero and that $B_1$, $B_2^a$, and $B_3$ are the same for all chosen energy eigenlevels $p_j^a$ from subspace $a$ and only a function of the entire probability distribution at a given instant of time.  Subtracting the equations of motion for the $i$th and $K$th energy eigenlevels results in
\begin{equation}
\frac{d}{dt}(\ln\frac{p_j^a}{n_j^a}-\ln\frac{p_k^a}{n_k^a})=-\frac{1}{\tau}(\ln\frac{p_j^a}{n_j^a}-\ln\frac{p_k^a}{n_k^a})+\frac{1}{\tau}\frac{B_3}{B_1}(\epsilon_j^a-\epsilon_k^a)
\end{equation}
Defining a new variable
\begin{equation}
W_{jk}=\frac{1}{\epsilon_j^a-\epsilon_k^a}(\ln\frac{p_j^a}{n_j^a}-\ln\frac{p_k^a}{n_k^a})
\end{equation}
the time evolution of $W_{jk}$ yields to the ordinary differential equation
\begin{equation}
\frac{dx}{dt}=-\frac{1}{\tau}x+\frac{1}{\tau}\frac{B_3}{B_1}
\end{equation}
If $p_j^a$ and $p_k^a$ are in the same subsystem for which the initial probability distribution is a canonical one, i.e., if
\begin{equation}
p_j^a(t=0) = \alpha^an_j^ae^{-\epsilon_j^a\beta^a},\quad p_k^a(t=0) = \alpha^an_k^ae^{-\epsilon_k^a\beta^a}
\end{equation}
then
\begin{equation}
W_{jk}(t=0)=\frac{1}{\epsilon_j^a-\epsilon_k^a}(\ln\frac{p_j^a}{n_j^a}-\ln\frac{p_k^a}{n_k^a})= -\beta^a
\end{equation}
For $\forall p_j^a,p_k^a$ in the same subsystem $a$, the time evolution of $W_{jk}$ yields to the same ordinary differential equation (ODE) with the same initial value, namely,
\begin{equation}
\frac{dx}{dt}=-\frac{1}{\tau}x+\frac{1}{\tau}\frac{B_3}{B_1},\ x=W_{jk}(t=0) =-\beta^a
\end{equation}
so that the solution of $W_{jk}$ is the same $ W_{jk}(t) =\beta^a(t)$. Therefore, the probability distribution in this subsystem maintains the canonical distribution with the parameter $\beta^a(t)$ given by
\begin{equation}
p_j^a(t) = \alpha^a(t)n_je^{-\epsilon_j^a\beta^a(t)}
\end{equation}
In addition, the temperature of the subsystem at time $t$ is defined by
\begin{equation}
T^a(t) = \frac{1}{k_b\beta^a(t)}
\end{equation}
 
Thus, for a system in a nonequilibrium state, the hypoequilibrium temperature for each subsystem is defined. This temperature can be the same or different from that of any other subsystem. If a system is in a $M$th-order hypoequilibrium state, it remains at least of order $M$ throughout as well as after the evolution, and the probability distribution of each subsystem remains canonical. 

\section{Evolution of intensive properties}
In this appendix, the evolutions of intensive properties are given for the system with probability and energy conservations. Eq. (\ref{EQM}) is reformulated such that
\begin{equation}
\frac{d}{dt}\ln\frac{p_j^K}{n_j^K}=\frac{1}{\tau}(-\ln\frac{p_j^K}{n_j^K}-\frac{A_2}{A_1}+\epsilon_j^a\frac{A_3}{A_1})
\end{equation}
Using Eqs. (\ref{HES}) and (\ref{34}) yields
\begin{equation}\label{B2}
\frac{d}{dt}(-\alpha^K(t)-\beta^K(t)\epsilon_i^K)=\frac{1}{\tau}(\alpha^K(t)+\beta^K(t)\epsilon_i^K-\alpha-\epsilon_j^K\beta)
\end{equation}
Subtracting the equations of motion for the $i$th and $j$th energy eigenlevels results in
\begin{eqnarray}
\frac{d}{dt}(-\beta^K(t)\epsilon_i^K+\beta^K(t)\epsilon_j^K)&=&\frac{1}{\tau}(\beta^K(t)\epsilon_i^K-\beta^K(t)\epsilon_j^K)\nonumber\\
&&-\frac{1}{\tau}\beta(\epsilon_i^K-\epsilon_j^K)
\end{eqnarray}
If $i\neq j$, dividing both sides by $(\epsilon_i^K-\epsilon_j^K)$ results in the evolution for $\beta^K$, namely,
\begin{equation}\label{B4}
\frac{d\beta^K}{dt}=-\frac{1}{\tau}(\beta^K-\beta)
\end{equation}
Finally, subtracting Eq. (\ref{B4}) from (\ref{B2}) gives the evolution for $\alpha^K$, i.e.,
\begin{equation}
\frac{d\alpha^K}{dt}=-\frac{1}{\tau}(\alpha^K-\alpha)\\
\end{equation}
\bibliography{Diffusion_paper_20160105}

\end{document}